\documentclass[useAMS,usenatbib]{mnras}
\usepackage{multirow}
\usepackage{graphicx,color}
\newcommand{\etal }{{et al.} }
\newcommand{\msun}{\thinspace M_\odot} 
 
\newcommand{\lsun}{\thinspace L_\odot}

\def\lesssim{\mathrel{\hbox{\rlap{\hbox{\lower4pt\hbox{$\sim$}}}\hbox{$<$}}}}
\def\gtrsim{\mathrel{\hbox{\rlap{\hbox{\lower4pt\hbox{$\sim$}}}\hbox{$>$}}}}
\newcommand{\cm}{\,{\rm cm}^{-3} } 
\newcommand{\km}{\,{\rm km\, s}^{-1}}

\newcommand{\dfrac}[2]{{\displaystyle \frac{#1}{#2}} }

\title[Massive Outflows II]{Massive Outflows Driven by Magnetic Effects II: Comparison with Observations}
\author[Y.~Matsushita  \etal]
  {Yuko Matsushita$^{1}$, Yuya Sakurai$^{2}$, Takashi Hosokawa$^{3}$, and Masahiro N. Machida$^{1}$\thanks{E-mail: yuko.matsushita.272@s.kyushu-u.ac.jp (YM),  machida.masahiro.018@m.kyushu-u.ac.jp (MNM)}  \\
$^{1}$ Department of Earth and Planetary Sciences, Faculty of Sciences, Kyushu University, Fukuoka 819-0395, Japan\\
$^{2}$ Department of Physics, The University of Tokyo, Tokyo 113-0033, Japan\\
$^{3}$Department of Physics, Graduate School of Science, Kyoto University, Sakyo-ku, Kyoto 606-8502, Japan
}

\begin{document}
\maketitle
\begin{abstract}
The driving mechanism of massive outflows observed in high-mass star-forming regions  is investigated using three-dimensional magnetohydrodynamics (MHD) and protostellar evolution calculations. 
In our previous paper, we showed that the mass outflow rate depends strongly on the mass accretion rate onto the circumstellar disk around a high-mass protostar, and massive outflows may be driven by the magnetic effect in high-mass star-forming cores. 
In the present study, in order to verify that the MHD disk wind is the primary driving mechanism of massive outflows, we quantitatively compare outflow properties obtained through simulations and observations. 
Since the outflows obtained through simulations are slightly younger than those obtained through observations, the time-integrated quantities of outflow mass, momentum, and kinetic energy are slightly smaller than those obtained through observations. 
On the other hand, time-derivative quantities of mass ejection rate, outflow momentum flux, and kinetic luminosity obtained through simulations are in very good agreement with those obtained through observations. This indicates that the MHD disk wind greatly contributes to the massive outflow driving from high-mass protostars, and the magnetic field might significantly control the high-mass star formation process. 
\end{abstract}

\section{Introduction}
\label{sec:intro}
Radio observations have confirmed many molecular outflows in various star-forming regions \citep[e.g.,][]{snell80,wu04,arce10}. 
Molecular outflows have a wide opening angle and a speed of $\gtrsim 1$--$10\km$ \citep{arce07,velusamy14}.
Since molecular outflows are ubiquitously observed around young stellar objects (YSOs), they are considered to universally appear in the star formation process \citep{lada85,pudritz86,shu87}.
The outflows are proof of mass accretion onto the circumstellar disk and/or protostar, because they are powered by the release of the gravitational energy of accreting matter \citep{konigl00}. 
In addition, molecular outflow ejects a large fraction of cloud mass and determines the star formation efficiency \citep{matzner00}.
Thus, the molecular outflow is essential to understanding the star formation process, and its driving mechanism should be clarified.

Although numerous molecular outflows have been observed in low-mass star-forming regions, which are located near the Sun, their driving mechanism has not been identified  \citep{arce07}. 
Controversial hypotheses on outflow driving have been proposed.
In a classical theory or model, the molecular outflow (hereinafter, outflow), which is usually observed at radio wavelengths, is entrained by another high-velocity component, namely, the optical jet (hereinafter, jet), which is usually observed at optical and near-infrared wavelengths.
We refer to this scenario as the entrainment hypothesis \citep{munt83,raga93,canto91,matzner99}. 
On the other hand, recent magnetohydrodynamics (MHD) simulations have shown that the low-velocity wide-angle outflow is driven {\it directly} by the circumstellar disk around a protostar \citep{tomisaka02,banerjee06,machida08,price12,tomida13,tomida15,lewis17}.
We refer to this scenario as the MHD disk wind hypothesis, in which the outflow is magneto-centrifugally driven by the rotationally supported disk \citep{blandford82}. 
Thus, two competing scenarios have been proposed as the outflow driving mechanism, which remains a topic of debate.

Recent high-resolution ALMA observations are beginning to provide proof for the MHD disk wind hypothesis.
\citet{bjerkeli16} showed that the molecular outflow is directly driven by the disk around low-mass YSO TMC-1A. 
Moreover, through ALMA observation, \citet{aso15} confirmed the existence of a Keplerian rotation disk perpendicular to the outflow axis around TMC-1A.  
In addition, over the past several years, several Keplerian rotation disks have been confirmed around (low-mass) Class 0 protostars, which are driving molecular outflows \citep{hara13,murillo13,codella14,lee14,ohashi14,sakai14,yen15}. Through ALMA observation, \citet{alves17} also provided strong evidence of the MHD disk wind around Class I protostars.
The results of recent high-resolution observations in low-mass star-forming regions are in good agreement with recent theoretical studies on low-mass star formation, in which molecular outflow is driven directly by the circumstellar (or Keplerian) disk, and is not entrained by a high-velocity jet \citep{inutsuka12}.

On the other hand, compared to the low-mass star formation process, high-mass star formation remains poorly understood. 
Since massive star-forming regions are located far from the Sun, it is difficult to directly observe high-mass protostars, disks, and their surrounding environments.
However, it is well known that high-mass protostars also drive (massive) outflows that have a size of $>0.1$\,pc \citep{beuther02}.
Thus, we can observe (or resolve) the outflows from high-mass protostars and might be able to derive some information about high-mass star formation from the observations of such outflows \citep{wu04}.

Very recently, using ALMA, \citet{hirota17} observed high-mass protostar Orion KL source I resolving the circumstellar disk and outflow and showed clear evidence of the MHD disk wind, in which the rotations of both disk and outflow were confidently identified \citep[see also][]{greenhill13,hirota16a,hirota16b}. 
Note that they could not find any sign of a high-velocity jet, and specifically excluded other outflow driving mechanisms, such as entrainment and radiation-driven mechanisms.
Although there are a small number of samples at this time, recent observations imply that high-mass protostars also drive outflow by the MHD disk wind mechanism.

Unlike low-mass star formation, radiation feedback might be considered for the outflow driving mechanism in high-mass star formation. 
However, in an analytical study, \citet{tanaka17} showed that the MHD disk wind is the primary feedback mechanism rather than feedback by radiation-driven and/or stellar winds, even for the high-mass star formation process. 
They also pointed out that over 90\% of outflow momentum is provided by the MHD disk wind in the mass range of $M_{\rm ps} \lesssim 100\msun$, where $M_{\rm ps}$ is the protostellar mass.
In addition to \citet{tanaka17}, a number of studies pointed out that the MHD disk wind mechanism is viable and important, even for the high-mass star formation process \citep{peters11,commercon11, hennebelle11, seifried12,kuiper15}.
Thus, it may be important to investigate high-mass star formation in terms of the MHD disk wind.

In the present paper, following our previous study \citep[][referred to hereinafter as Paper I]{matsushita17}, we investigate the massive outflow driven by high-mass protostars. 
In Paper I,  we calculated the evolution of massive clouds and outflows parameterizing the mass accretion rate according to the core accretion scenario \citep{tan13,tan14} and showed that massive outflow driven by the MHD disk wind mechanism may appear in the massive star formation process when the parent clouds for high-mass star formation have strong magnetic fields. 
In the present study, using calculation data taken from Paper I, we estimate the protostellar luminosity in order to directly compare outflow properties derived from simulations with those obtained through observations.
Then, we confirm whether massive outflows from high-mass protostars are indeed driven by the MHD disk wind mechanism.

The remainder of the present paper is organized as follows. Section 2 describes the numerical settings of the proposed model, and the calculation of protostellar evolution is described in Section 3.  Section 4 presents the simulation results.
We compares the simulation results with observations in Section 5, and discuss caveats in comparing outflow properties between simulations and observations in Section 6.
Finally, the paper is summarized in Section 7.

\section{Numerical Settings and Models}
\label{sec:model}

\begin{table*}
\begin{center}
\begin{tabular}{c|ccccccccccc|ccccccc} \hline
{\footnotesize Model} & $M_{\rm cl} $ & $R_{\rm cl}$ & $B_0$ & $\Omega_0$ & \multirow{2}{*}{$\alpha_0$} & \multirow{2}{*}{$\beta_0$} & \multirow{2}{*}{$\gamma_0$} & \multirow{2}{*}{$\mu$}  & $\dot{M}$ \\
 &  {\scriptsize [$\msun$]} & {\scriptsize [pc]} & [$\mu$\,G] & {\scriptsize [$10^{-14}$\, s$^{-1}$]} &  &  &  & & {\scriptsize [$\msun$\, yr$^{-1}$] }  \\
\hline
A   & 32  & \multirow{6}{*}{0.28} &  23 & 3.3 & 0.5 & \multirow{6}{*}{0.02} & \multirow{6}{*}{0.2} & \multirow{6}{*}{2} & $8.2 \times 10^{-5}$ \\
B   & 77  & \multirow{4}{*}{0.28} &  56 &                 5.1     & 0.2 &  &  &   & $2.2 \times 10^{-4}$ \\
C   & 192 &                       &  140 &                8.0      & 0.08 &  &  & & $8.6 \times 10^{-4}$ \\
D   & 385 &                       &  280 &                11      & 0.04 &  &  &  & $2.0 \times 10^{-3}$ \\
E   & 771 &                       &  560 &                16      & 0.02 &  &  &  & $6.2 \times 10^{-3}$ \\
F   & 1,542&                       &  1,120&                23      & 0.01&  &  &   & $1.7 \times 10^{-2}$ \\
\hline
\end{tabular}
\end{center}
\caption{
Model parameters.
Column 1 lists the model name. 
Columns 2--5 list the cloud mass $M_{\rm cl}$, cloud radius $R_{\rm cl}$, magnetic field strength $B_0$, and angular velocity $\Omega_0$, respectively. 
Columns 6, 7, and 8 list the ratios of thermal $\alpha_0$, rotational $\beta_0$, and magnetic energy $\gamma_0$, respectively, to the gravitational energy of the initial cloud core.
Column 9 lists the initial mass-to-flux ratio normalised by the critical value $\mu$.
Column 10 lists the average mass accretion rate at the end of simulation.
}
\label{table:1} 
\end{table*}  

As described in Section 1, we used simulation data taken from Paper I. 
Since the numerical settings and model parameters are described in Paper I, we explain them only briefly here.

Using a resistive MHD nested grid code \citep[for details, see][]{machida04,machida05a,machida05b,machida07,machida10,machida13} that covers a spatial scale of 0.8\,AU--$9.1\times10^5$\,AU, we calculated the evolution of star-forming clouds until the protostellar mass reaches $\sim2$--$100\msun$ with a sink, in which the sink threshold density $n_{\rm sink}$ and accretion radius $r_{\rm sink}$ were set to $n_{\rm sink}=10^{13}\cm$ and $r_{\rm sink}=1$\,AU, respectively.
In Paper I, we prepared fiducial models A--F and models CE1-CE4, CW1, CW2, EW1, and EW2 (see Table~1 of Paper I). Since models CE1-CE4, CW1, CW2, EW1, and EW2 were calculated to investigate only the dependence of resistivity (CE1-CE4) and magnetic field strength (CW1, CW2, EW1, and EW2), in the present paper, we use fiducial models A--F.

The model parameters are listed in Table~\ref{table:1}. 
The initial clouds have a Bonner-Ebert density profile with a central density $\rho_{\rm c,0}=3.8\times10^{-19}$\,g$\cm$ and an isothermal temperature $T_{\rm iso}=40$\,K. 
To promote contraction, we enhance the cloud density by a factor of $f=1.4-67$, where $f$ is the density enhancement factor and changes the ratio of thermal to gravitational energy of the initial clouds $\alpha_0$ (Paper I).   
The initial clouds have the same radius of $R_{\rm cl}=$0.28 pc and different masses in the range of $M_{\rm cl}=32-1542\msun$ (Table~\ref{table:1}). 
A uniform magnetic field  and rigid rotation  are set as the initial state, in which the magnetic field 
is assumed to be parallel to the rotation axis.
The magnetic field strengths are adjusted to have $\mu=2$, where $\mu$ is the mass-to-flux ratio of the initial clouds normalized by the critical value $(2\pi G^{1/2})^{-1}$.
The angular velocities are also chosen  to have $\beta_0=0.02$, where $\beta_0$ is the ratio of  rotational to gravitational energy of the initial clouds. 
In Table~\ref{table:1}, $B_0$ and $\Omega_0$ of each model is listed. 
The mean mass accretion rate for each model is also described in Table~\ref{table:1} (for details, see Paper I). The mass accretion rate is controlled by the ratio of thermal to gravitational energy $\alpha_0$ of the prestellar cloud, and characterizes the models: High-mass stars form with high mass accretion rates, and low-mass stars form with relatively low mass accretion rates. A more detailed description of the numerical settings and model parameters can be found in Paper I. 

\section{Protostellar Evolution and Luminosity}
\label{sec:luminosity}
The protostellar evolution is followed by numerically solving the interior structure as a post-process, i.e., with accretion histories pre-calculated by the 3D MHD simulations as in \citet{machida13}.
We use the {\it STELLAR} numerical code \citep{yorke08,hosokawa13,sakurai15}, which uses the Henyey method to construct stellar models. STELLAR differs from the code used in \citet{machida13}, which assumes the spherical accretion flow directly hits the stellar surface \citep[e.g.,][]{stahler80,hosokawa09}. In contrast, STELLAR assumes the disk accretion with a control parameter $\eta$ ($0 \leq \eta \leq 1$), where part of the accretion luminosity $L_{\rm *,acc} = \eta L_{\rm acc}$ is assumed to be temporarily deposited to the stellar interior. This approach has been widely used to model the thermal property of the accreting gas \citep[i.e., the specific entropy, e.g.,][]{siess97}, which is unknown in the stellar evolution calculations. 
For all of the examined cases, we adopt $\eta = 0.01$.
It should be noted that we confirmed that the choice of $\eta$ does not significantly affect the results, in which the 
accretion luminosity can vary by a factor of six at the maximum in the range of $0.01\le \eta \le 1$ \citep{baraffe12,kuiper13,baraffe17}.  
We also confirmed that the resulting characteristic properties of high-mass protostars, such as the large stellar radius, are qualitatively and quantitatively the same when the different code, which is described above, is adopted  \citep[e.g.,][and Sec.~4.2.1 below]{hosokawa10}. 
We start the evolutionary calculations from an arbitrary tiny initial model with $0.01~M_\odot$, which roughly corresponds to  the initial mass of the protostar \citep{larson69, masunaga00}.

\section{Overview of Simulation}
\label{sec:results}
Through three-dimensional simulations, we calculated the evolution of each cloud listed in Table~\ref{table:1} and previously showed the detailed structure of outflows and disks in Paper I. 
Thus, in this section, we simply provide an overview of the results.

\begin{figure*}
\includegraphics[width=160mm]{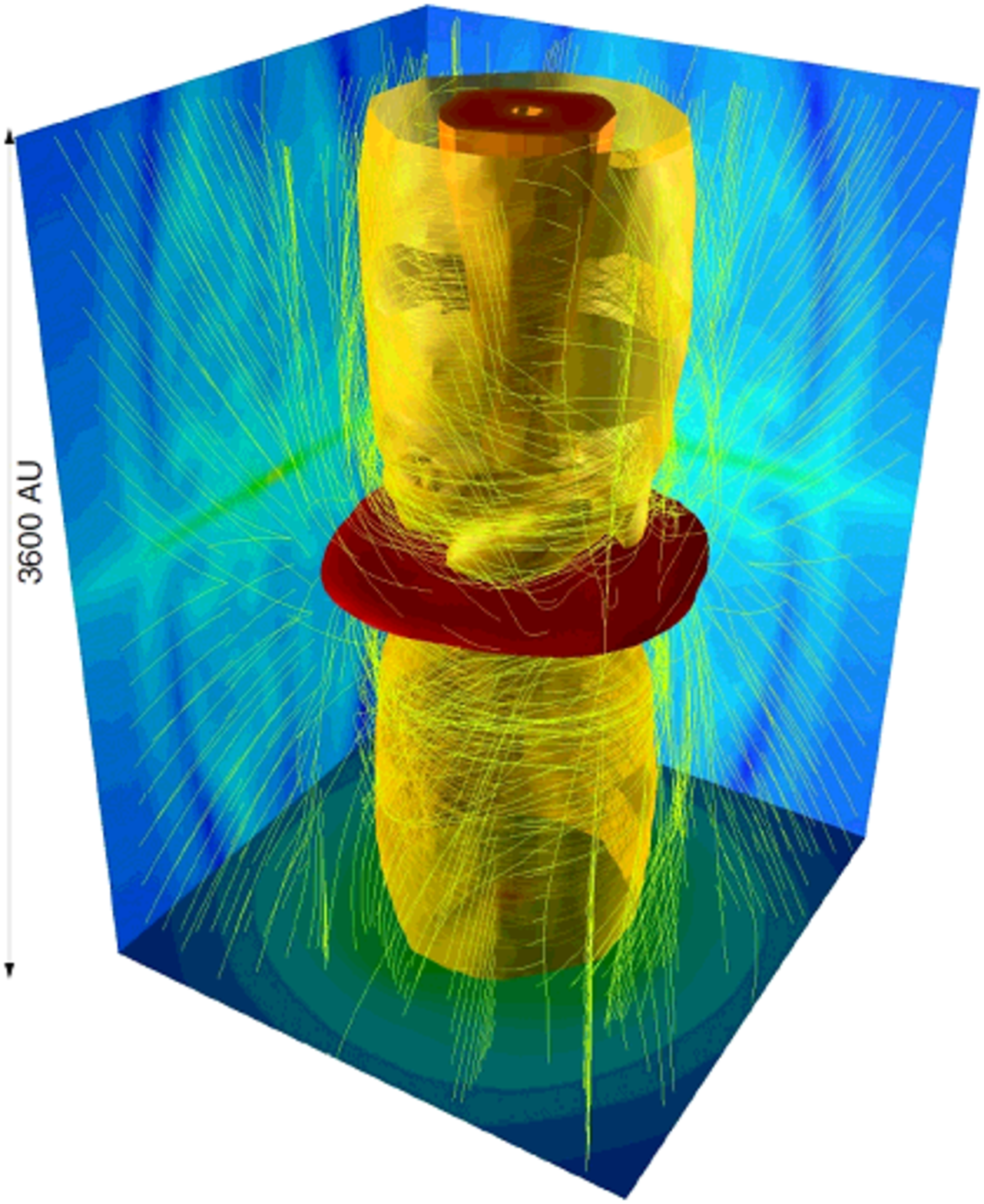}
\caption{
Three-dimensional view of model E when the protostellar mass reaches $M_{\rm ps}=10\msun$ at $t_{\rm ps}=2{,}417.3$\,yr, where $t_{\rm ps}$ is the elapsed time after protostar formation.
Yellow and orange surfaces are iso-velocities of $v_r=1$ and 10 $\km$, respectively, and correspond to the outflow driving region.
The central red region is an iso-density surface of $\rho=6.0\times10^{-15}{\rm g}\cm$ and corresponds to the pseudo disk.
The magnetic field lines are indicated by the yellow lines. 
The density distribution on the $x=0$, $y=0$, and $z=0$ planes are projected onto the surface of each wall.  
The spatial scale is shown to the left of the panel.
}
\label{fig:1}
\end{figure*}

Figure~\ref{fig:1} shows a three-dimensional view of outflow and magnetic field lines for model E. The figure indicates that outflow is driven by the disk-like structure at the cloud centre, and  magnetic field lines are strongly twisted inside the outflowing region (or inside the yellow surface). 
The well-collimated structure of outflow is caused by the hoop stress. 
The strongly twisted magnetic field lines and the well-collimated structure of outflow are typical features of the MHD disk wind \citep{kudoh98,konigl00,tomisaka02}.

\begin{figure*}
\includegraphics[width=160mm]{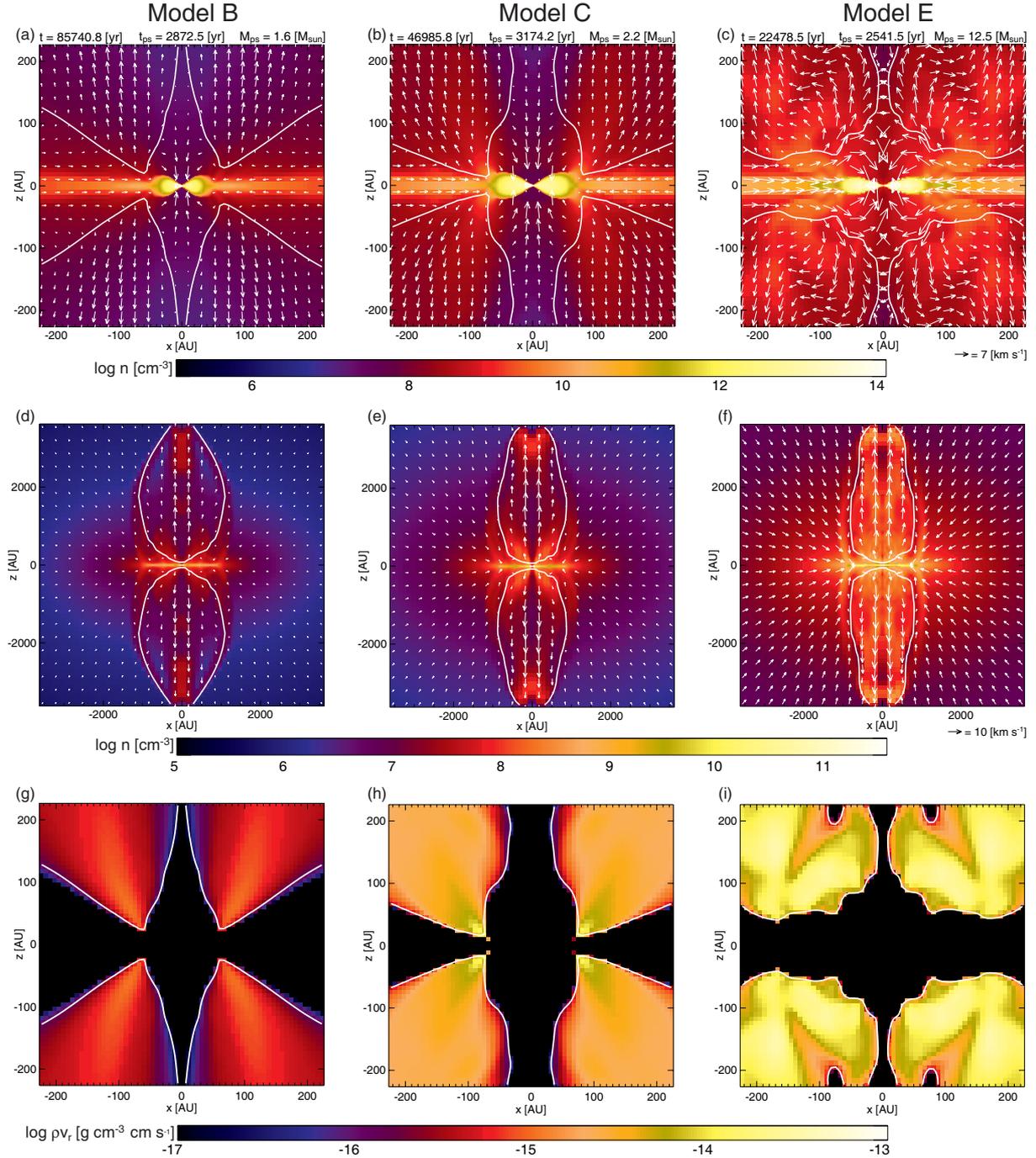}
\caption{
Top and middle panels: Density (colour) and velocity (arrows) distributions for models ({\it a} and {\it d}) B, ({\it b} and {\it e}) C, and ({\it c} and {\it f}) E when the outflow of each model reaches $\sim3{,}000$\,AU. 
The box scale is different in the ({\it a}-{\it c}) top and ({\it d}-{\it f}) middle panels. 
Bottom panels: outflow momentum $\rho \, v_r$ at each point for models ({\it g}) B, ({\it h}) C, and ({\it i}) E at the same epoch as the top and middle panels. 
The box scale is the same as in the top panels. 
The white contours in each panel indicate the boundaries between the infalling region and the outflowing region, inside which the gas is outflowing from the central region. 
The elapsed time after the calculation starts ($t$) and that after protostar formation ($t_{\rm ps}$) are listed above each top panel. 
The protostellar mass ($M_{\rm ps}$) is also described in each top panel.  
}
\label{fig:2}
\end{figure*}

In order to compare the difference in strength of outflows between models, the density and velocity distributions on the $y=0$ plane for models B, C, and E  are plotted in the top and middle panels of Figure~\ref{fig:2}. 
As described in Table~\ref{table:1}, models B, C, and E have average mass accretion rates of $\dot{M} = 2.2 \times 10^{-4} \msun$\,yr$^{-1}$, $8.6\times10^{-4}\msun$\,yr$^{-1}$, and $6.2\times 10^{-3}\msun$\,yr$^{-1}$, respectively.
The velocity and density of the outflow in the high-mass accretion model (model E) are larger than those in the low-mass accretion models (models B and C).
However, there is no significant difference in structure in large-scale outflows (Fig.~\ref{fig:2} middle panels), in which the outflow in each model has a well-collimated structure.

The outflow momentum ($\rho\, v_r$) at each point is plotted in the bottom panels of Figure~\ref{fig:2}. 
These panels show that the outflow momentum in the model with a higher mass accretion rate is relatively large, as explained in Paper I. 
As shown in Figures~\ref{fig:1} and \ref{fig:2}, although the structure of the outflow does not significantly differ between models, the outflow strength or momentum differs considerably among models. 
As described in Paper I, the difference in outflow properties among the models is attributed to the accretion rate onto the circumstellar disk, because the outflow is powered by the release of the gravitational energy of the accreting matter, which is proportional to the mass accretion rate $\dot{M}_{\rm acc}$. 
Therefore, the difference in outflow physical quantities should also be attributed to the mass accretion rate, which is related to the initial cloud stability $\alpha_0$ (for details, see Paper I).

\section{Comparison with Observations}
\label{sec:comparison}
In order to further validate the MHD disk wind as the (primary) driving mechanism of massive outflow in high-mass star-forming regions, in this section, we compare the simulation results with observations.

\subsection{Outflow Length and Protostellar Luminosity}
\label{sec:length}
\begin{figure*}
\includegraphics[width=160mm]{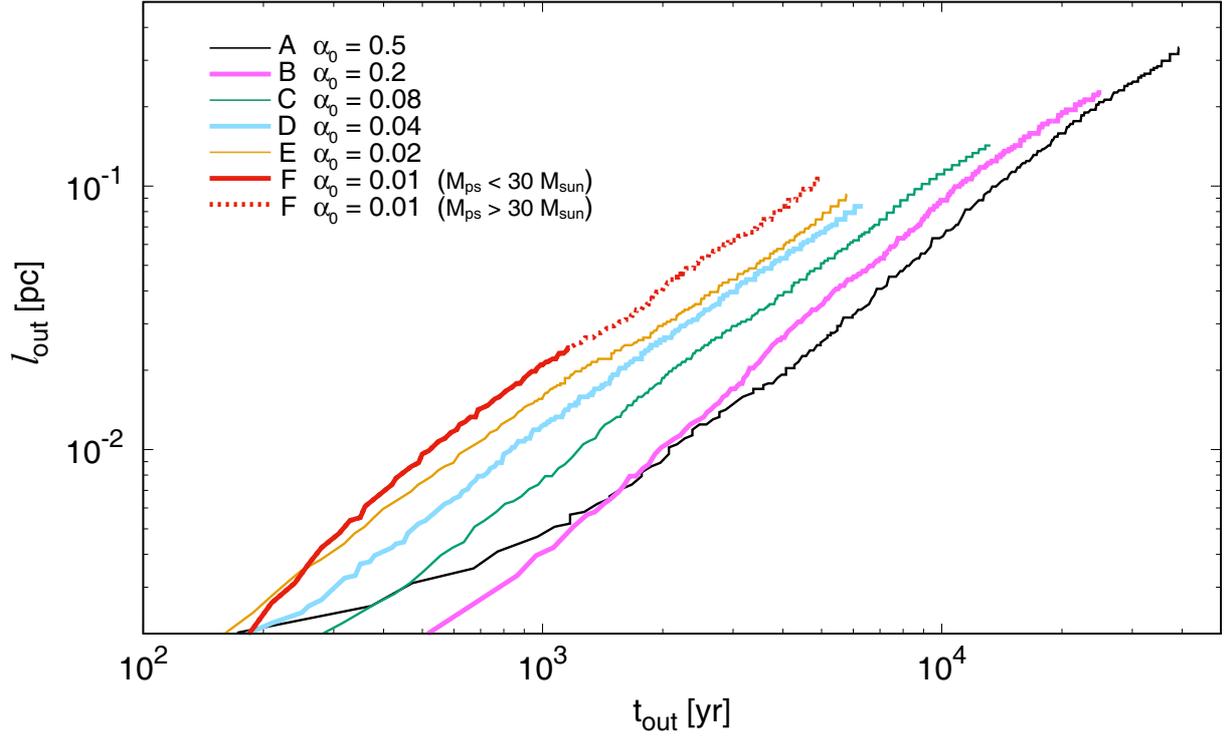}
\caption{
Outflow length for all models plotted with respect to the elapsed time after outflow emerges. 
For reference, the outflow length for model F after the protostellar mass reaches $M_{\rm ps}=30\msun$ is also plotted as the red dotted line.
}
\label{fig:3}
\end{figure*}

Figure~\ref{fig:3} shows the evolution of the outflow length with respect to the elapsed time after the outflow emerges. 
The outflows reach $\sim 0.1$--$0.3$\,pc by the end of the calculation.
Note that we  plotted the calculation results before the protostellar mass reaches $M_{\rm ps}\sim20$--$30\msun$ because the radiative effects cannot be ignored for $M_{\rm ps}\gtrsim 20$--$30\msun$, as discussed in Paper I \citep[see also][]{yorke02,krumholz07,kuiper10,kuiper15,kuiper16}.
However, \citet{tanaka17} showed that the MHD disk wind is the primary feedback mechanism, even when the protostellar mass reaches $M_{\rm ps} \sim 100\msun$.
Thus, for reference, for model F only, we added the simulation data after the protostellar mass exceeds $M_{\rm ps}=30\msun$.

In the following, we compare outflow properties derived from simulations with observations of massive outflows \citep{wu04,beuther02,zhang05,villiers14,maud15}.
The outflow lengths listed in \citet{beuther02}, \citet{wu04}, \citet{maud15} and \citet{villiers14} are in the range of $\sim 0.1-2$\,pc and comparable or slightly larger than those derived from our simulations (Fig.~\ref{fig:3}). 
The large-scale outflows are preferentially observed, while it takes a considerably large computational time to reach the outflow length of $\sim 2$\,pc in simulations. 
Thus, it should be noted that, in some cases, we may compare relatively younger outflows in simulations with relatively older outflows in observations.
Therefore, in comparison with outflows in simulations, some outflows in observations have relatively large amounts of mass, momentum and kinetic energy  (\S\ref{sec:integrated}). 
Instead,  time-derivative outflow quantities in observations are in well agreement with those in simulations (\S\ref{sec:der}).

\begin{figure*}
\includegraphics[width=160mm]{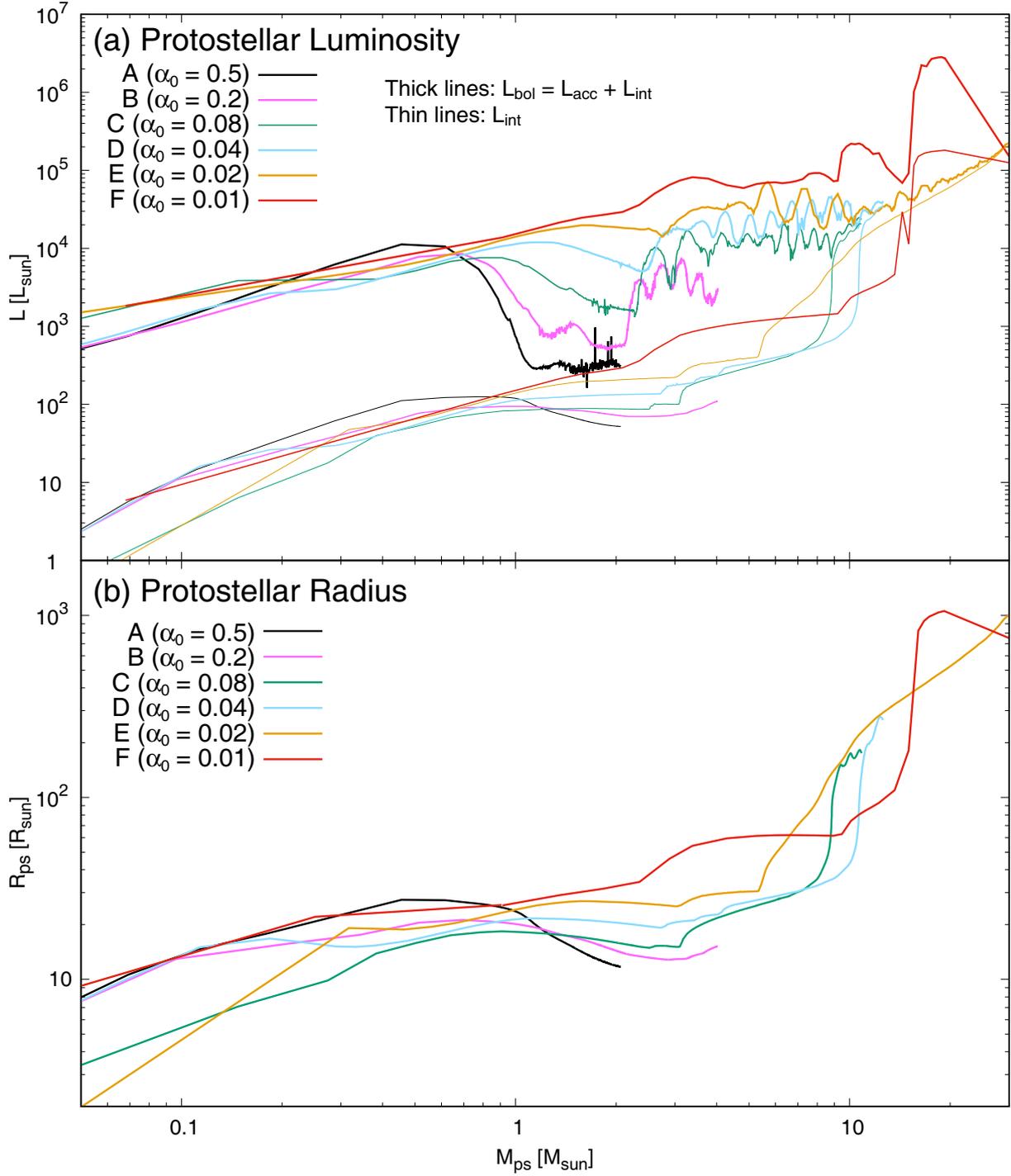}
\caption{
Protostellar luminosity (upper panel) and radius (lower panel) for each model plotted with respect to the protostellar mass. 
In the upper panel, the thick and thin lines correspond to the total bolometric ($L_{\rm bol}=L_{\rm acc}+L_{\rm int}$; thick lines) and internal ($L_{\rm int}$; thin lines) luminosities of each protostar, where $L_{\rm acc}$ is the stellar accretion luminosity.
The relation between the parameter $\alpha_0$ and the mass accretion rate $\dot{M}$, which is proportional to the accretion luminosity,  can be referred in Table~\ref{table:1} (see also, Paper I).
}
\label{fig:4}
\end{figure*}

As described in Section 1, we cannot directly observe high-mass protostars. 
Thus, the bolometric luminosity, which is considered to be emitted by protostars, is used as the measure of protostellar mass in the observations.  
Note that the bolometric luminosity acquired through observations is integrated over the entire massive star-forming cloud that is illuminated by a high-mass protostar.
In order to more precisely compare the simulations and observations, we calculated the protostellar luminosity $L_{\rm bol}$ ($=L_{\rm acc}+L_{\rm int}$) for each model (Section 3) and plotted the obtained values in the upper panel of Figure~\ref{fig:4}, 
in which the stellar internal luminosity $L_{\rm int}$ is also plotted, where $L_{\rm acc}$ the stellar accretion luminosity.
The figure indicates that the accretion luminosity dominates the internal luminosity by the end of the calculation in all models. The oscillation of bolometric luminosities in Figure~\ref{fig:4} is attributed to the time variability of the mass accretion. 
For a high mass accretion rate (see the last column of Table~\ref{table:1}), the surface density of the circumstellar disk increases for a short duration and causes gravitational instability, which develops a spiral or non-axisymmetric structure \citep{toomre64} and induces time-variable mass accretion onto the protostar, as discussed in Paper I.

Although the protostellar luminosity depends on the mass accretion rate, it is in the range of $10^2\lesssim L/\lsun \lesssim 10^6$ in Figure~\ref{fig:4}.
The protostellar luminosities derived from simulations are roughly comparable to the observations (see below). 
Thus, from the viewpoint of protostellar luminosity, we can compare massive outflows at almost the same age between the simulations and observations.
However, it should be note that there exist some exceptions in observation, in which some massive protostars in observations have somewhat larger bolometric luminosities than simulations.
As also described above, they are expected to correspond to more evolve (or older) protostars, and have larger outflow quantities (\S\ref{sec:integrated}).

The lower panel of Figure~\ref{fig:4} shows that, in all cases, the high-mass accreting protostar has a large radius of from few tens of to one hundred  solar radii. This agrees with previous calculations, which generally adopted constant accretion rates for simplicity \citep[e.g.,][]{hosokawa09,hosokawa10}. 
Note that such a large radius is linked to the resulting protostellar luminosity because the accretion luminosity $L_{\rm acc} \propto R_{\rm ps}^{-1}$ currently dominates the bolometric luminosity, as described above.

\subsection{Time-integrated Outflow Quantities vs.\ Bolometric Luminosity}
\label{sec:integrated}
\begin{figure*}
\includegraphics[width=160mm]{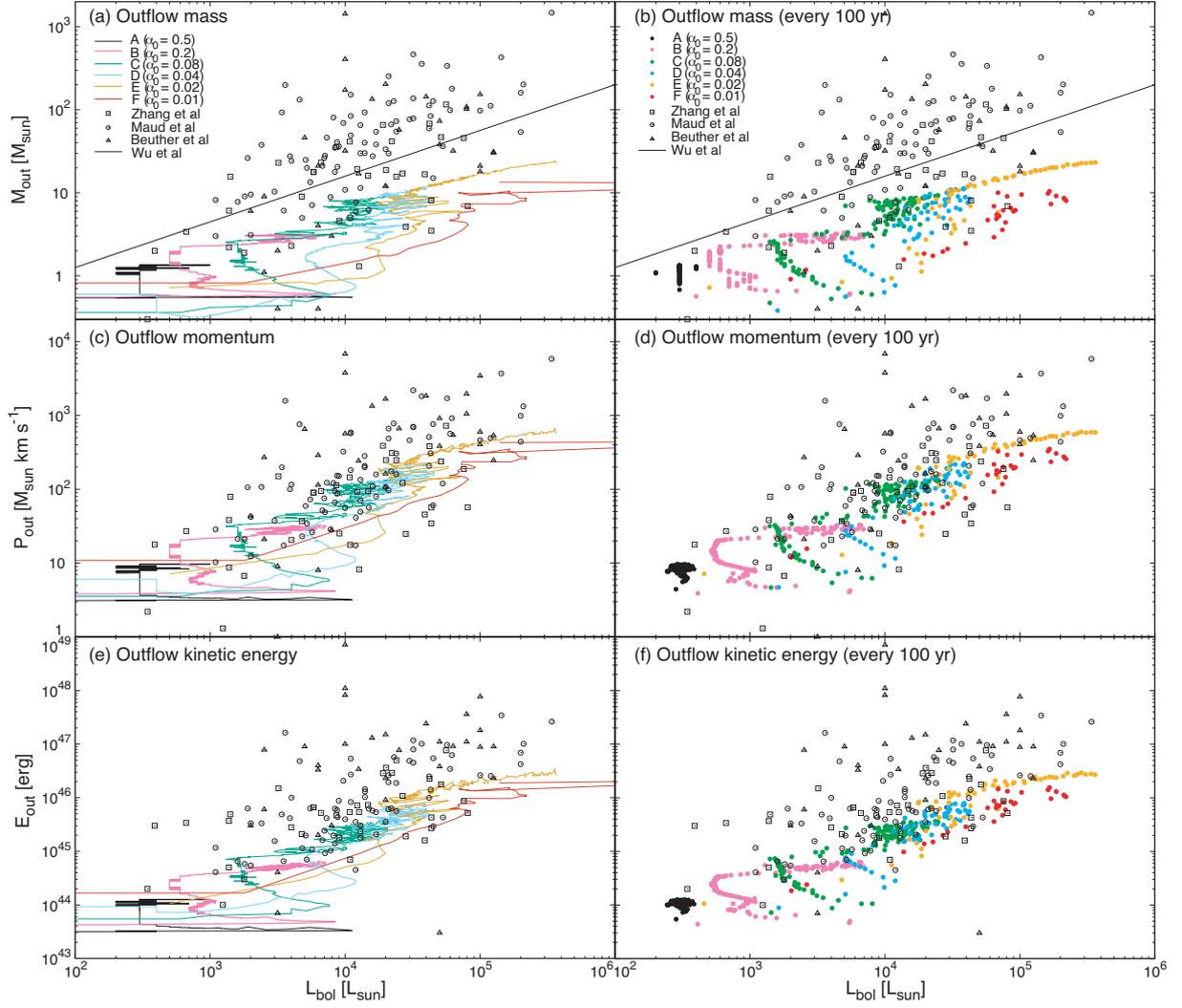}
\caption{
({\it a} and {\it b}) Outflow mass, ({\it c} and {\it d}) momentum, and  ({\it e} and {\it f}) kinetic energy for simulation models A--F (coloured lines in the left panels and coloured dots in the right panels) with respect to the protostellar (or bolometric) luminosity. 
The coloured points in the right-hand panel are the simulation results for every 100 yr. 
Observations taken from tables in \citet[][squares]{zhang14}, \citet[][circles]{maud15}, and \citet[][triangles]{beuther02} are also plotted as black symbols. 
The solid lines in panels ({\it a}) and ({\it b}) are derived from large samples in the observations \citep{wu04}.
}
\label{fig:5}
\end{figure*}

We compared the outflow physical properties in simulations with those in observations with respect to the bolometric luminosity in Figure~\ref{fig:5}, in which observations taken from tables of  \citet{zhang14}, \citet{maud15}, and \citet{beuther02} are plotted as symbols. Simulation results for every 100 years are plotted as coloured symbols in the right-hand panels of Figure~\ref{fig:5}, and the clustered regions of colour symbols are predicted to be preferentially observed if massive outflows are driven by the MHD disk wind mechanism. 
The outflow mass, momentum, and kinetic energy are plotted in Figure~\ref{fig:5}. 
Since they are time-integrated quantities, the outflow mass, momentum, and kinetic energy might depend on the evolutional stage of the protostar or outflow. 
We discuss this in greater detail in \S\ref{sec:der} and \S\ref{sec:discussion}.

The masses of outflows are plotted in Figures~\ref{fig:5}{\it a} and {\it b}, in which the black solid lines correspond to the fitting formula given by \citet{wu04} and are given by 
\begin{equation}
{\rm log}\, (M_{\rm out}/M_{\rm sun}) = -1.04 + 0.56\, {\rm log}\, (L_{\rm bol}/L_{\rm sun}).
\end{equation}
The figures indicate that, as a whole, the outflow masses derived from simulations are slightly less massive than those obtained from observations. 
The outflow masses in simulations are in the range of $0.1 \lesssim M_{\rm out}/\msun \lesssim 20$, whereas those in observations are in the range of $0.1 \lesssim M_{\rm out}/\msun \lesssim 10^3$. 
Thus, very massive outflows ($M_{\rm out}\gtrsim 20\msun$) did not appear in simulations. 
Note that we plotted the outflow mass before the protostar reaches $M_{\rm ps}=30\msun$ because the radiative feedback can be ignored for $M_{\rm ps}<30\msun$ \citep{krumholz07,kuiper10,kuiper15,kuiper16}.
Moreover, we confirmed that the outflow mass reaches $M_{\rm out}\sim100\msun$ when the protostellar mass becomes $M_{\rm ps} \sim100\msun$. 

The same tendency can be seen in the outflow momentum (Figs.~\ref{fig:5}{\it c} and {\it d}). 
The outflow momenta in the simulation are in the range of $5\lesssim P_{\rm out}/ (\msun \, {\rm km}\, {\rm s}^{-1}) \lesssim 5\times 10^2$, whereas those in observations are in the range of $1\lesssim P_{\rm out}/ (\msun \, {\rm km}\, {\rm s}^{-1}) \lesssim 10^4$.
Thus, although some outflow momenta derived from simulations are in good agreement with the observations,  the observed values are occasionally larger than those derived from simulations. 

The outflow kinetic energies are plotted in Figures~\ref{fig:5}{\it e} and {\it f}. 
As indicated by the outflow mass and momentum, some outflow kinetic energies are in agreement with the observations, whereas others are smaller than the observations. 
As described in \S\ref{sec:length}, the outflows derived from simulations are considered to be somewhat younger than observed outflows. 
Since the physical quantities in Figure~\ref{fig:5} are time-integrated quantities, it may be natural that outflow properties in some simulation results are smaller than the observations.

\subsection{Time-derivative Outflow Quantities vs.\ Bolometric Luminosity}
\label{sec:der}
\begin{figure*}
\includegraphics[width=160mm]{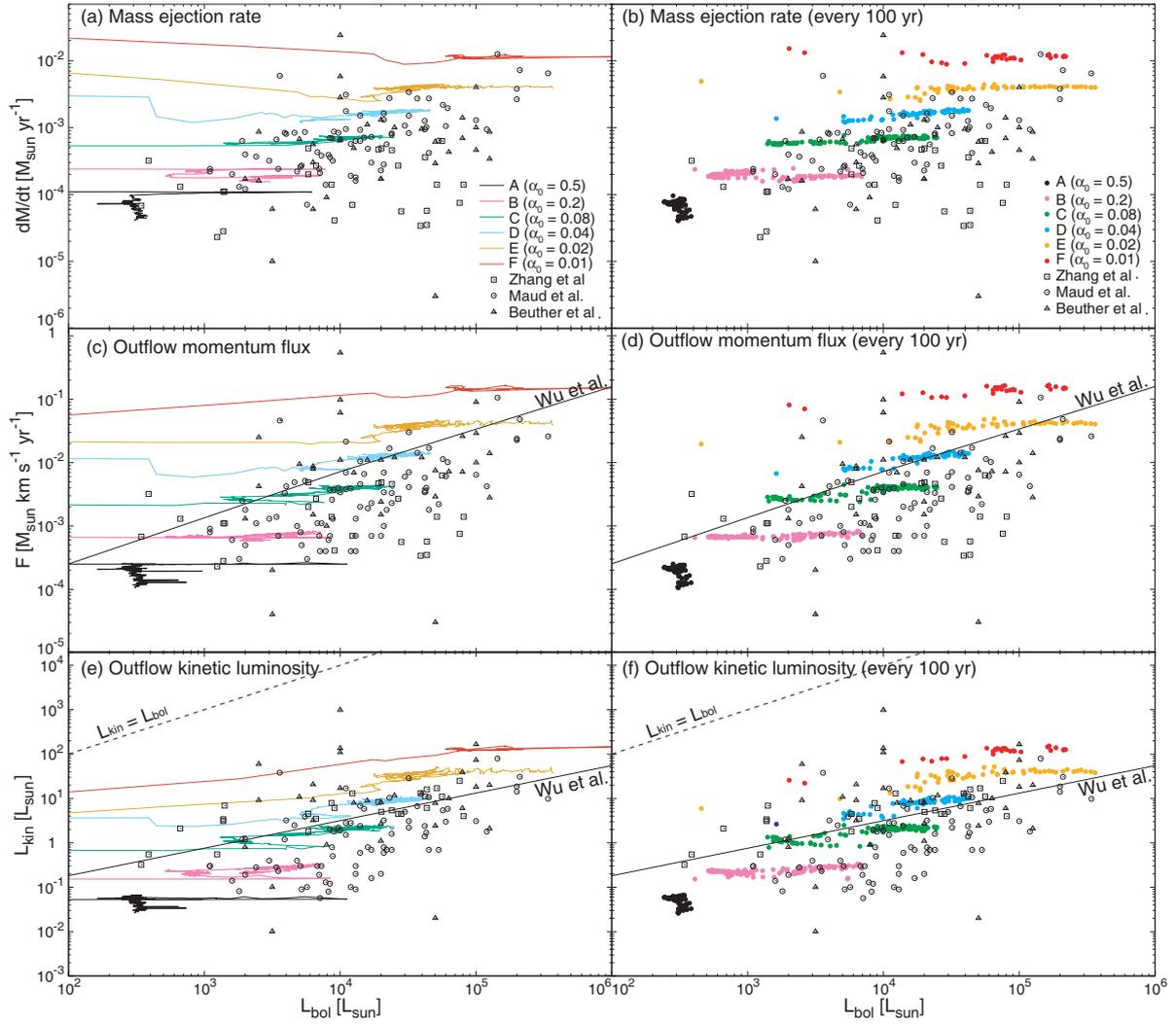}
\caption{
 ({\it a} and {\it b}) Mass ejection rate, ({\it c} and {\it d}) outflow momentum flux, and ({\it e} and {\it f}) kinetic luminosity for simulation models A--F (coloured lines in the left-hand panels and coloured dots in the right-hand panels) plotted with respect to the protostellar (or bolometric) luminosity. 
The coloured points in the right-hand panel are the simulation results for every 100 yr.
Observations taken from tables in \citet[][squares]{zhang14}, \citet[][circles]{maud15}, and \citet[][triangles]{beuther02} are also plotted as the black symbols. 
The solid lines in panels ({\it a}) through ({\it d}) are derived from large samples in observations \citep{wu04}.
The relation $L_{\rm kin}= L_{\rm bol}$ is plotted in panels ({\it e}) and ({\it f}). 
}
\label{fig:6}
\end{figure*}

Assuming steady outflow, the outflow mass, momentum, and kinetic energy should depend on the evolutional stage of outflow or the elapsed time after outflow emerges, because they are time-integrated quantities. 
On the other hand, time-derivative quantities might be better for comparing simulations and observations and for identifying the outflow driving mechanism. 
The outflow mass ejection rate, momentum flux, and kinetic luminosity, which are time-derivative quantities, for each model are plotted with respect to the bolometric luminosity in Figure~\ref{fig:6}, in which the outflow mass, momentum, and kinetic energy are each divided by the elapsed time after the outflow emerges as follows:
\begin{eqnarray}
\dot{M} = M_{\rm out}/ t_{\rm out}, \\
F = P_{\rm out}/t_{\rm out}, \\
L_{\rm kin} = E_{\rm out}/t_{\rm out},
\end{eqnarray}
where $t_{\rm out}$ is the elapsed time after the outflow emerges.

Figure~\ref{fig:6}{\it a} shows that the mass ejection rate for each simulation model does not depend significantly on the bolometric luminosity, where each model has almost the same mass ejection rate. 
Depending on the mass accretion rate (or $\alpha_0$), the mass ejection rates are in the range of $10^{-4} \lesssim \dot{M}/(\msun \, {\rm yr}^{-1})\lesssim 10^{-2}$.
Although observations shows some small mass ejection rates of $\dot{M}< 10^{-4}\msun$\,yr$^{-1}$, many of them correspond well to simulations.
Figure~\ref{fig:6}{\it b} shows that the distribution of mass ejection rates in the simulations are in good agreement with those in the observations. 
However, some observational data are plotted in the lower right region, where no simulation data exist. 
In such a region, more evolved high-mass protostars just before or after the main accretion phase ends may be observed, because the mass ejection rate is low and the bolometric luminosity is high. 
Note that we could not calculate the outflow evolution by the end of the mass accretion stage.
Moreover, we clearly showed that the mass ejection rate is proportional to the mass accretion rate in Paper I.

In Figures~\ref{fig:6}{\it c} and {\it d}, the outflow momentum fluxes derived from the simulations are roughly in agreement with the observations. 
The outflow momentum fluxes derived through simulation are somewhat larger than those of the observations. 
The fitting formula derived in \citet{wu04} 
\begin{equation}
{\rm log}\, (F/(\msun {\rm km}\, {\rm s}^{-1}\, {\rm yr}^{-1})  )  = -4.92 +  0.648\, {\rm log}\,(L_{\rm bol}/L_{\rm sun})
\end{equation} 
is plotted in these figures. 
The simulation results appear to agree well with the fitting formula.
The momentum flux has the dimensions of force and is considered to be important in determining the outflow driving mechanism \citep[e.g.][]{cabrit92, bontemps96,hatchell07}. 
Figures~\ref{fig:6}{\it c} and {\it d} imply that massive outflows observed in high-mass star-forming regions may be driven by the MHD disk wind mechanism because the outflow momentum fluxes in the MHD simulations are sufficient to explain the observations. 
As seen in the bottom panels of Figure~\ref{fig:2}, the outflow momentum (flux) increases as the mass accretion rate onto the protostar increases. 
The high mass accretion rate is necessary to form a high-mass star.
Thus, Figures~\ref{fig:2}, \ref{fig:6}{\it e} and {\it f} are the evidence for MHD disc winds responsible for massive outflows observed around high-mass protostars.

As shown in Figures~\ref{fig:6}{\it e} and {\it f}, the outflow kinetic luminosities derived from the simulation are also comparable to the observations. 
The fitting formula in these figures is taken from \citet{wu04} and is as follows:
\begin{equation}
{\rm log}\, (L_{\rm kin}/L_{\rm sun}) = -1.98 + 0.62\, {\rm log}\,(L_{\rm bol}/L_{\rm sun}). 
\end{equation} 
$L_{\rm kin}=L_{\rm bol}$ is also plotted in the figure. 
The outflow kinetic luminosities are significantly smaller than the protostellar bolometric luminosities, indicating that the massive outflows are not driven by the radiative effect \citep{wu04}.


\section{Discussion}
\label{sec:discussion}
We compared the simulations and observations in \S\ref{sec:comparison}.
Overall, the simulation results are in good agreement with the observations. 
Thus, the massive outflow observed in high-mass star-forming regions can be explained by the MHD disk wind. 
However, there is a discrepancy between the simulations and observations, and there are caveats concerning the calculations.  
In this section, after discussing the possible reasons for this discrepancy, we discuss the abovementioned caveats. 

\subsection{Age Difference between Simulations and Observations}
In \S\ref{sec:comparison}, we first compared the outflow properties of the mass, momentum, and kinetic energy in the simulations with those in  the observations, and they showed {\it better} agreement with observations.  
However, some outflows in the observations have larger physical quantities than the simulations because of the {\it slightly} different evolutionary stage of the outflows, as described in \S\ref{sec:results}.
Next, we compared the time-derivative quantities of outflows (mass ejection rate, outflow momentum flux, and kinetic luminosity) between the simulations and observations, and {\it very good} agreement was found between them.
Since it is expected that the time-derivative quantities do not significantly depend on the evolutionary stage, they are essential for determining the outflow driving mechanism. 
Thus, our results strengthen the hypothesis that the massive outflows are driven primarily by the MHD disk wind.

However, there is discrepancy in the (dynamical) timescale of outflow between simulations and observations.
We showed that outflow ages are {\it substantially} different between the simulations and observations in Paper I (see Fig.~10 of Paper I).
The elapsed time of outflow in the simulations was $\sim10^4$\,yr, whereas, in the observations, the outflow dynamical timescale, which is defined as  
\begin{equation}
t_{\rm dyn,obs}=\ell_{\rm out}/v_{\rm out}, 
\label{eq:dynobs}
\end{equation}
where  $\ell_{\rm out}$ and $v_{\rm out}$ are the observed outflow length and typical (or averaged)  outflow velocity, is $t_{\rm dyn,obs} \sim10^5$\,yr. 
Thus, the difference in the outflow timescale between the simulations and observations is approximately one order of magnitude. 
In contrast, although some outflows in the observations have larger integrated physical quantities than those in the simulations, many of the simulation results are comparable to the observations, as shown in Figure~\ref{fig:5}, in which the physical quantities are plotted with respect to the bolometric luminosity. These results might imply that the outflow dynamical timescale in the observations is not very accurate.

In summary, some outflows in the simulations appear to be younger than those in the observations, as described in \S\ref{sec:results} and Figures~\ref{fig:3} and \ref{fig:5}, whereas some outflows in simulations appear to have approximately the same outflow length (Fig.~\ref{fig:3}) and protostellar luminosity (Fig.~\ref{fig:5}), indicating that the protostellar and outflow ages between simulations and observations do not differ greatly. 
On the other hand, although outflows have approximately the same physical quantities, there are gaps between the outflow ages between simulations and observations, as shown in Figure 10 of Paper I.

Although we can determine the elapsed time after outflow emerges without uncertainty in simulations, there may exist uncertainty when estimating the outflow velocity and length in observations. 
Even when the outflow inclination angle and outflow length are correctly determined, it is difficult to determine the outflow velocity. 
In simulations, outflows show time variability, in which the outflow (typical or maximum) velocity also changes with time. 
In addition, since massive star-forming regions are located very far from the Sun, the beam size or spatial resolution of telescopes is not always sufficient to resolve the high-velocity component, which would be embedded in a dense gas.

\begin{figure*}
\includegraphics[width=160mm]{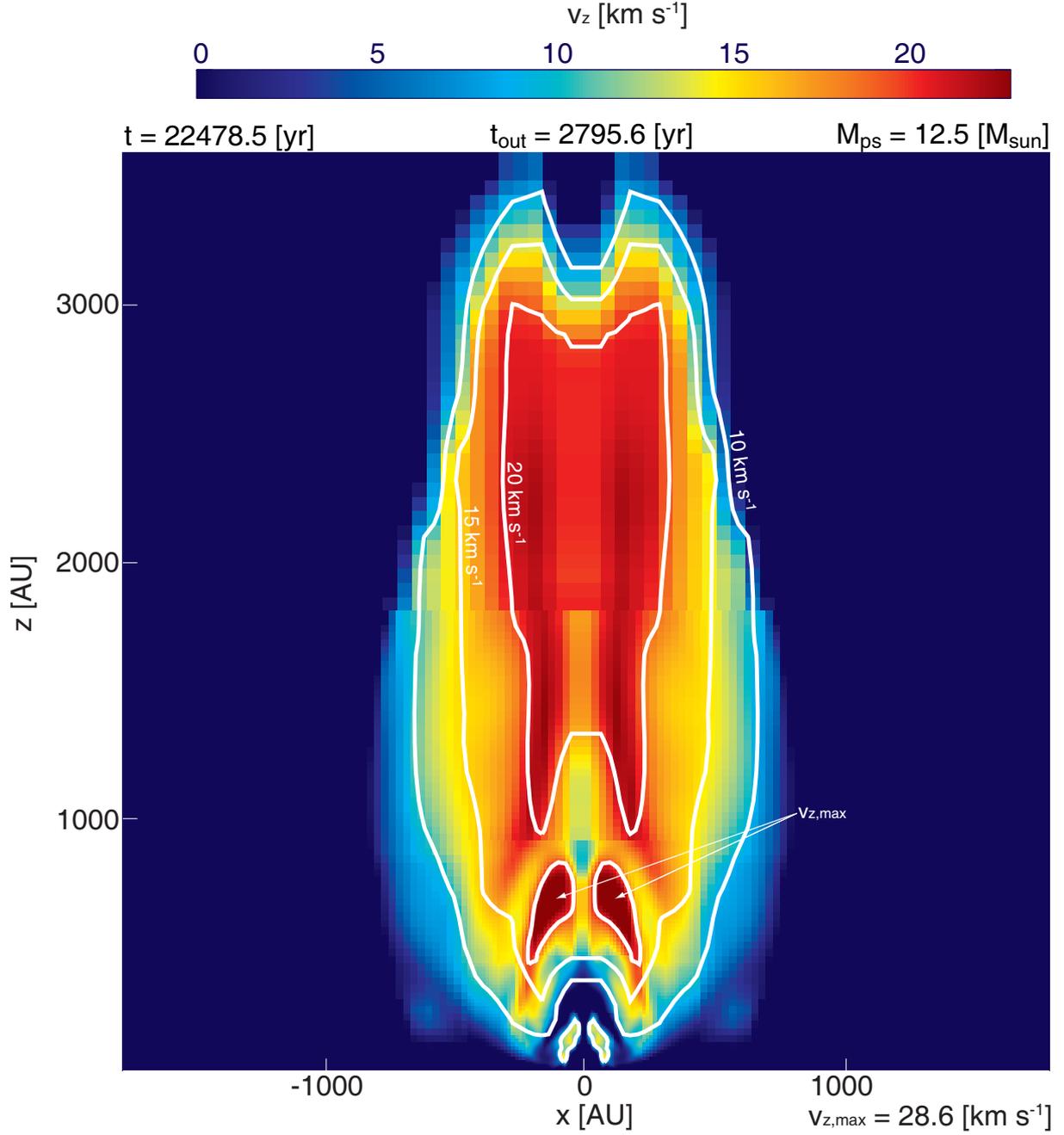}
\caption{
Outflow velocity $v_z$ (colour) on the $y=0$ plane for model E at the same epoch as in Figure~\ref{fig:2}, in which only the $z>0$ region is plotted.    
The elapsed time after the calculation starts ($t$) and that after outflow emerges ($t_{\rm ps}$) are listed above the panel. 
The protostellar mass ($M_{\rm ps}$) is also described. 
The labelled velocity contours are plotted as the white lines. 
The positions of the outflow maximum velocity are indicated by arrows. 
}
\label{fig:7}
\end{figure*}

Figure~\ref{fig:7} shows the distribution of the outflow velocity on the $y=0$ plane for model E at the same epoch as in Figure~\ref{fig:2} and indicates that the high-velocity component is embedded in the low-velocity component.
In addition, the outflow has a maximum velocity of $\sim 30$\,km\,s$^{-1}$ near its root, whereas the propagation velocity is approximately 6 km\,s$^{-1}$ ($\ell_{\rm out}/t_{\rm out}\simeq 3{,}500\,{\rm AU}/2{,}796$\,yr). 
Thus, the difference between the maximum velocity and the propagation velocity is approximately five-fold. 
In the observations, the dynamical timescale is estimated by a snapshot of outflows.
However, Figure~\ref{fig:7} indicates that it is difficult to identify the dynamical timescale from a snapshot in the observations.

In order to further investigate the dynamical timescale derived from simulations, we calculated it as  
$t_{\rm dyn, \rho^2}=\ell_{\rm out}/v_{\rm out,\rho^2}$, where we define $v_{\rm out,\rho^2}$ as
\begin{equation}
v_{\rm out,\rho^2} = \dfrac{\int_{{v_r} > v_{\rm cri}}\, v_{\rm r} \, \rho^2\, dV }{\int_{v_r > v_{ \rm cri}}\, \rho^2\, dV},
\label{eq:tdyn}
\end{equation}
where $v_{\rm cri}=0.5\km$ is adopted (for details, see Paper I).
The massive outflows are observed in the shock-excited molecular line emission from carbon monoxide \citep{beuther02,zhang05,villiers14,maud15}.
Thus, we weighted the massive outflows by squared density in order to derive the outflow dynamical timescale, because the intensity of such a line emission is proportional to the squared density. 
For reference, we also calculated the density-weighted velocity,
\begin{equation}
v_{\rm out,\rho} = \dfrac{\int_{{v_r} > v_{\rm cri}}\, v_{\rm r} \, \rho\, dV }{\int_{v_r > v_{ \rm cri}}\, \rho\, dV},
\label{eq:tdyn}
\end{equation}
and derived the outflow dynamical timescale as $t_{\rm dyn, \rho}=\ell_{\rm out}/v_{\rm out,\rho}$.

\begin{figure*}
\includegraphics[width=160mm]{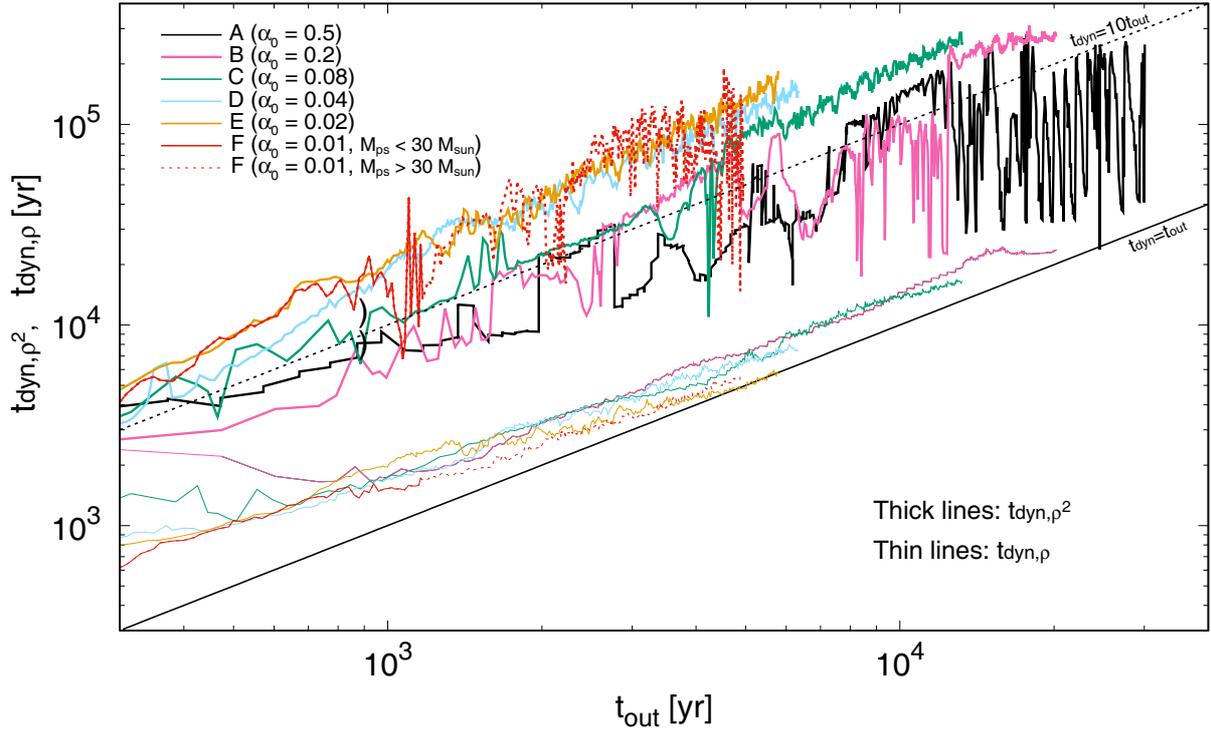}
\caption{
Outflow dynamical timescale $t_{\rm dyn,\rho^2}$ (thick lines) and $t_{\rm dyn,\rho}$ (thin lines) plotted with respect to the elapsed time after the outflow emerges $t_{\rm out}$.
The dynamical timescale after the protostellar mass reaches $M_{\rm ps}=30\msun$ for model F is also plotted as the red dotted lines.
}
\label{fig:8}
\end{figure*}

Figure~\ref{fig:8} compares the outflow dynamical timescale $t_{\rm dyn,\rho^2}$ and $t_{\rm dyn,\rho}$ with the elapsed time $t_{\rm out}$ after outflow emerges.
The figure indicates that the dynamical timescale oscillates because of the time variability of the outflow. 
In addition, based on the figure, the dynamical timescale $t_{\rm dyn,\rho^2}$ is approximately 10-times longer than the elapsed time $t_{\rm out}$. 
The dynamical timescale $t_{\rm dyn, \rho}$ is also slightly longer than the elapsed time $t_{\rm out}$. 
This is because the outflow length is determined by a relatively high-velocity component, which is embedded in a dense gas and is difficult to observe, as shown in Figures~\ref{fig:1} and \ref{fig:7}. 
On the other hand, the outflow has a massive low-velocity component, which encloses the high-velocity component (Figs.~\ref{fig:1} and \ref{fig:2}) and is easily observed. 
The mass of the high-velocity component is small \citep{machida14}, so the outflow physical quantities are considered to be determined by the (massive) low-velocity component. 

The observation may underestimate the maximum or typical outflow velocity. 
In such a case, since the observed outflow is young, the discrepancy in the outflow timescale might be resolved. 
Note that the dynamical timescale in Figure~\ref{fig:8} is a simple estimation. 
This figure simply indicates the possibility of the observed dynamical timescale being systematically longer than the elapsed time after outflow emergence. 
More realistically, we need a strict treatment to compare the dynamical timescale between simulations and observations, in which molecular chemistry and line emission should be considered. 
However, such treatment is far beyond the scope of the present study.

\subsection{Spatial Resolution in Simulations}
Next, we discuss the caveats in our calculations. 
As discussed previously in Paper I, we used a sink method and did not spatially cover the region near the protostar. 
Thus, the high-velocity jet ($\gtrsim 100$\,km\,s$^{-1}$), which may be driven near the protostar \citep{machida08}, could not be resolved. 
Since the mass of the high-velocity component is expected to be very small, the high-velocity component (or jet) does not significantly contribute to the estimation of the total (low-velocity) outflow mass, momentum, and kinetic energy \citep{machida14}.
However, the sink may influence the estimation of the dynamical timescale, because the high-velocity component might determine the outflow length if it exists. 
This is also related to the difference in the outflow timescale between simulations and observations.

Observations of \citet{villiers14} revealed that the maximum velocity of massive outflows is in the range of $2\, {\rm km}\,{\rm s}^{-1} \lesssim v_{\rm out,max}\lesssim 15\, {\rm km}\,{\rm s}^{-1}$, which is comparable to our simulations (Fig.~\ref{fig:7}).
Thus, no high-velocity component ($\gtrsim 100$\,km s$^{-1}$) appears in the observations. 
In addition, \citet{hirota17} pointed out that they could find no evidence of the high-velocity component (or jet) in their observation of Orion KL source I.  
Thus, we have no clear evidence of the high-velocity jet, which is sometimes observed around the low-mass protostar, driven from the high-mass protostar. 
Recent ALMA observations do not support the entrainment hypothesis \citep{bjerkeli16,alves17,hirota17}, in which the high-velocity jet greatly contributes to driving the low-velocity outflow, as described in \S\ref{sec:intro}.
On the other hand, in the disk wind hypothesis, the high-velocity jet and outflow are driven from different places (or different radii) of the circumstellar disk \citep{tomisaka02,banerjee06,machida08,tomida13}, in which the jet does not contribute to the outflow driving. 
In such a case, we can safety compare outflows between simulations and observations without resolving the region near the protostar where the high-velocity jet may appear.

The purpose of the present study was to investigate high-mass star formation from the viewpoint of massive outflows. 
As described above, although there is a minor discrepancy between the simulations and observations, the simulations appear to be able to explain the observations. 
In order to more precisely investigate the high-mass star formation and massive outflow, further time-integration and higher spatial resolution are necessary in simulations. 
Moreover, higher spatial resolution and more samples are required in observations.

\section{Summary} 
We studied the driving mechanism of massive outflows observed in high-mass star-forming regions. 
In Paper I, we related the mass accretion rate to the mass ejection (outflow) rate and showed that massive outflows may be driven by the MHD disk wind mechanism in highly gravitationally unstable cores that have a high mass accretion rate onto the circumstellar disk. 
In the present paper, in order to investigate the contribution of the MHD disk wind in massive outflows, we compared outflow physical quantities between simulations and observations.  
As a result, time-integrated quantities of outflow mass, momentum, and kinetic energy in the simulation were comparable to or slightly smaller than those of the observations. 
The outflow lengths in the simulations were slightly smaller than those in the observations.
It takes a long time to calculate sufficiently evolved outflows in simulation, whereas more evolved outflows should be preferentially observed in high-mass star-forming regions. 
Thus, it is natural that such time-integrated quantities in simulations are smaller than those in observations.

Next, we compared time-derivative quantities of mass ejection rate, outflow momentum flux, and outflow kinetic luminosity between simulations and observations and confirmed that the simulation results were in good agreement with the observations.
This indicates that the MHD disk wind greatly contributes to the driving massive outflows, although the effects of radiation, stellar wind, etc., cannot be completely ignored.
Therefore, the magnetic field greatly influences the massive star formation, and the basic mechanism of massive star formation is the same as in low-mass star formation, in which the star grows by the mass accretion onto the disk, and excess angular momentum is transported by magnetic effects, such as magnetic braking and outflow.

\section*{Acknowledgements}

The present study has benefited greatly from discussions with T.~Hirota, K.~Tomida, K.~Motogi, M.~Sekiya, and T.~Tsuribe. 
We are very grateful to an anonymous reviewer for a number of useful suggestions and comments. 
This research used the computational resources of the HPCI system provided by the Cyber Science Center at Tohoku University, the Cybermedia Center at Osaka University, and the Earth Simulator at JAMSTEC through the HPCI System Research Project (Project ID: hp150092, hp160079, hp170047).
Simulations reported in this paper were also performed by 2017 Koubo Kadai on Earth Simulator (NEC SX-ACE) at JAMSTEC. 
The present study was supported in part by JSPS KAKENHI Grant Numbers JP15J08816, JP15K05032, JP17H02869, JP17H06360, JP17K05387, and JP16H05996 and by the Advanced Leading Graduate Course for Photon Science.

\clearpage
\clearpage

\end{document}